\documentclass[conference]{IEEEtran}
\IEEEoverridecommandlockouts
\usepackage{cite}
\usepackage{amsmath,amssymb,amsfonts}
\usepackage{algorithmic}
\usepackage{graphicx}
\usepackage{textcomp}
\usepackage{booktabs}
\usepackage{xcolor}
\usepackage{tabularx}
\usepackage{caption}
\usepackage{subcaption}
\usepackage{xurl}
\newcommand{\projectName}{ROIBIN-SZ}

\def\BibTeX{{\rm B\kern-.05em{\sc i\kern-.025em b}\kern-.08em
    T\kern-.1667em\lower.7ex\hbox{E}\kern-.125emX}}
\begin{document}

\setlength{\floatsep}{1.2em}
\setlength{\textfloatsep}{1em}
\setlength{\intextsep}{0.2em}
\setlength{\belowcaptionskip}{0.2em}

\title{\projectName: Fast and Science-Preserving Compression for Serial Crystallography
%\thanks{This research was supported by the Exascale Computing Project (17-SC-20-SC), a collaborative effort of the U.S. Department of Energy Office of Science and the National Nuclear Security Administration. $\dagger$ these authors contributed equally to the drafting of the paper}
}

%\author{}%for blind review
\author{\IEEEauthorblockN{Robert Underwood $\dagger$}
\IEEEauthorblockA{\textit{Mathematics and Computer Science Division} \\
\textit{Argonne National Laboratory}\\
Lemont, USA \\
runderwood@anl.gov}
\and
\IEEEauthorblockN{Chunhong Chuck Yoon $\dagger$}
\IEEEauthorblockA{\textit{Information Systems Spec} \\
\textit{Stanford National Accelerator Laboratory}\\
Stanford, USA \\
yoon82@SLAC.Stanford.EDU}
\and
\IEEEauthorblockN{Ali Gok}
\IEEEauthorblockA{\textit{Cerebras Systems} \\
San Francisco, USA}
\and
\IEEEauthorblockN{Sheng Di}
\IEEEauthorblockA{\textit{Mathematics and Computer Science Division} \\
\textit{Argonne National Laboratory}\\
Lemont, USA \\
sdi1@anl.gov}
\and
\IEEEauthorblockN{Franck Cappello}
\IEEEauthorblockA{\textit{Mathematics and Computer Science Division} \\
\textit{Argonne National Laboratory}\\
Lemont, USA \\
cappello@mcs.anl.gov}
}

\maketitle

\begin{abstract}
Crystallography is the leading technique to study atomic structures of proteins and produces enormous volumes of information that can place strains on the storage and data transfer capabilities of synchrotron and free-electron laser light sources.
Lossy compression has been identified as a possible means to cope with the growing data volumes; however, prior approaches have not produced sufficient quality at a sufficient rate to meet scientific needs.
This paper presents Region Of Interest BINning with SZ lossy compression (\projectName{}) a novel, parallel, and accelerated compression scheme that separates  the dynamically selected preservation of key regions with lossy compression of background information.
We perform and present an extensive evaluation of the performance and quality results made by the co-design of this compression scheme.
We can achieve up to a 196$\times$ and 46.44$\times$ compression ratio on lysozyme and selenobiotinyl-streptavidin while preserving the data sufficiently to reconstruct the structure at bandwidths and scales that approach the needs of the upcoming light sources.
\end{abstract}


\section{Introduction}
%state of the world
Upcoming synchrotron and free-electron laser light sources such as Argonne National Laboratory's Advanced Photon Source (APS) and SLAC National Accelerator Laboratory's Linac Coherent Light Source-II High Energy (LCLS-II-HE) facilities will provide greater insights into the fundamental structures of materials, chemistry, and biology than what is possible with current systems.
They are expected to accomplish this through both increased spatial/temporal resolution and increased data rates.
Together they are expected to produce enormous volumes of data at line rates of approximately 1 TB/s, requiring the use of high-performance computing to process the data stream.

%problem with the state of the world
The corresponding parallel file system and network fabrics used to store the data generally have substantially lower capacity and data transfer bandwidth than needed. 
%at approximately 100 GB/s sustained bandwidth and comparatively limited storage capacity.
Even if 1,000 PB of storage were 
available---more than all storage at the Argonne Leadership Computing Facility  
combined---these light sources would exhaust the available capacity in less than 12 days of continuous operation.
This situation presents critical bottlenecks for scientists planning to study fundamental structures using these light sources.
Although non-hit rejection and lossless compression technology exist, these methods will be insufficient with the coming light sources.
The reason is  that light source data contains substantial background noise---even after gain correction and preprocessing---so that they are  difficult to compress with lossless compressors that rely on identifying patterns in the data.
Lossy compression has been recognized as a  promising solution to this problem. However, even leading state-of-the-art lossy compressors such as SZ introduce too much data distortion in the critical regions of the data to accomplish important tasks such as constructing electron density models at expected compression ratios. 
%that are critical to the scientific mission of these light sources at compression ratios that will have a meaningful impact on the volume of data produced.

%solution to the problem
We propose a specialized,  parallel lossy compression scheme, called ROIBIN-SZ (Region Of Interest BINning with SZ lossy compression),  for data produced by these instruments that achieves high compression ratios and bandwidth to scale to the needs of these systems while ensuring the scientific integrity of the results from the decompressed data.

Several challenging issues have to be resolved when developing an optimized lossy compression method for the scientific datasets generated in crystallography. First, the crystallography data generated by light source facilities are  noisy, so that it is nontrivial to obtain a high compression ratio even using the best existing state-of-the-art lossy compressors such as SZ \cite{taoSignificantlyImprovingLossy2017,sz16,sz3} and ZFP \cite{lindstromFixedRateCompressedFloatingPoint2014}. Second, investigating the impact of lossy compression on the post hoc analysis in crystallography is nontrivial. Specifically, a comprehensive study may cover different datasets generated by different detectors on different crystal structures. This end-to-end verification of the science results also calls for a series of sophisticated tools \cite{CollaborativeComputationalProject1994,McCoy2007} and analysis methods \cite{Yoon2018,White2016a}. Third,  many existing state-of-the-art lossy compression algorithms  have been proposed, each with distinct design principles and thus potentially largely different compression performances and qualities. \footnote{LibPressio, a lossy compression adaptor library, features at least 30 distinct methods of compression}. 

While stages of lossy compression techniques can be similar (i.e., quantization, huffman encoding, dictionary encoding may appear similar) \cite{cappelloUseCasesLossy2019}, designing a lossy compressor that achieves high compression ratios and preserves scientific integrity at high bandwidth---especially given the specific quality and performance challenges described above--- is  challenging because  of the large solution space and relatively small set of feasible solutions that meet all requirements.  Validating that these requirements have been met requires careful co-design with domain scientists and attention to hardware.

In this work we make the following contributions.
\begin{enumerate}
    \item We develop a flexible hardware and software co-design, ROIBIN-SZ, that can adapt automatically to new hardware and easily adapt to diverse compression methods.
    \item We develop a number of performance optimization strategies that preform at scale.
    \item We perform quality assessments on serial crystallography datasets collected at LCLS, based on our developed compression solution  ROIBIN-SZ, as compared with multiple related state-of-the-arts. Experiments verified no degradation of scientific outcome based on standard crystallography metrics.
    \item With faithful/acceptable fidelity on reconstructing electron densities, our suggested approach achieves an improvement of up to 17$\times$ in compression ratio over the leading lossy compressors and a gain of 329$\times$ over lossless compressors on lysozyme or 25.65$\times$ and 72$\times$ on selenobiotinyl-streptavidin.
\end{enumerate}

The rest of the paper is organized as follows. In Section \ref{sec:background} we introduce the background of the research. In Section \ref{sec:relate} we discuss the state-of-the-art compression methods used in crystallography data. In Section \ref{sec:design} we present an overview of our design ROIBIN-SZ. In Section \ref{sec:performanceopt} we describe our performance optimization strategies. In Section \ref{sec:experimentalsetup} we describe the experimental setup for compression hardware, software, and configuration. In Section \ref{sec:qualityassesment} we describe how the quality of the reconstructions were assessed. In Section \ref{sec:performanceassesment} we  evaluate our solution by comparing it with multiple other state-of-the-art solutions. In Section \ref{sec:conclusions} we provide a discussion and concluding remarks. .

%roadmap
\section{Background}\label{sec:background}
Crystallography is a method to obtain atomic structure of a protein. Conventional crystallography rotates a single large protein crystal in the x-ray beam to systematically collect a complete 3D diffraction volume that can be used to Fourier synthesize the electron density of the protein. More recently, radiation damage and difficulty in growing large crystals have led to the popularity of serial crystallography which uses the concept of ``diffraction-before-destruction.'' Instead of a single large crystal, tiny micro-sized crystals are injected into the x-ray beam in a serial manner, resulting in a single diffraction pattern being captured for a single crystal at random orientation before the onset of radiation damage. A typical LCLS experiment collects around a million diffraction patterns on a 4-megapixel detector. Each diffraction pattern is noisy and unique because of the random nature of the x-ray fluence/spectra and crystal size, orientation, and position in the beam. Therefore, correlations in time and space are hard to exploit for compression algorithms.

Because of the random nature of the injection process, some fraction of the dataset does not contain diffraction from a crystal, and typically images containing less than 15 peaks are considered not useful and can be discarded, a process  called ``non-hit rejection'' (NHR).  
NHR is geared towards modest precision and high recall as not to reject potential hits that may ultimately index---a process that orients the molecule.
While NHR is not a part of the contributions of our work, we mention it because it provides peak locations that we use in our technique.
The focus of this work is on compression techniques that can be used, in addition to the savings afforded by NHR.
Although NHR helps greatly reduce storage requirements, we do not include NHR contributions in any compression ratio figures reported unless otherwise specified.
Peak finding and NHR are an active area of research and beyond this scope of this paper \cite{clabbers2021myd88,boutet2012high}.

\begin{figure}
    \centering
    \includegraphics[width=.8\columnwidth]{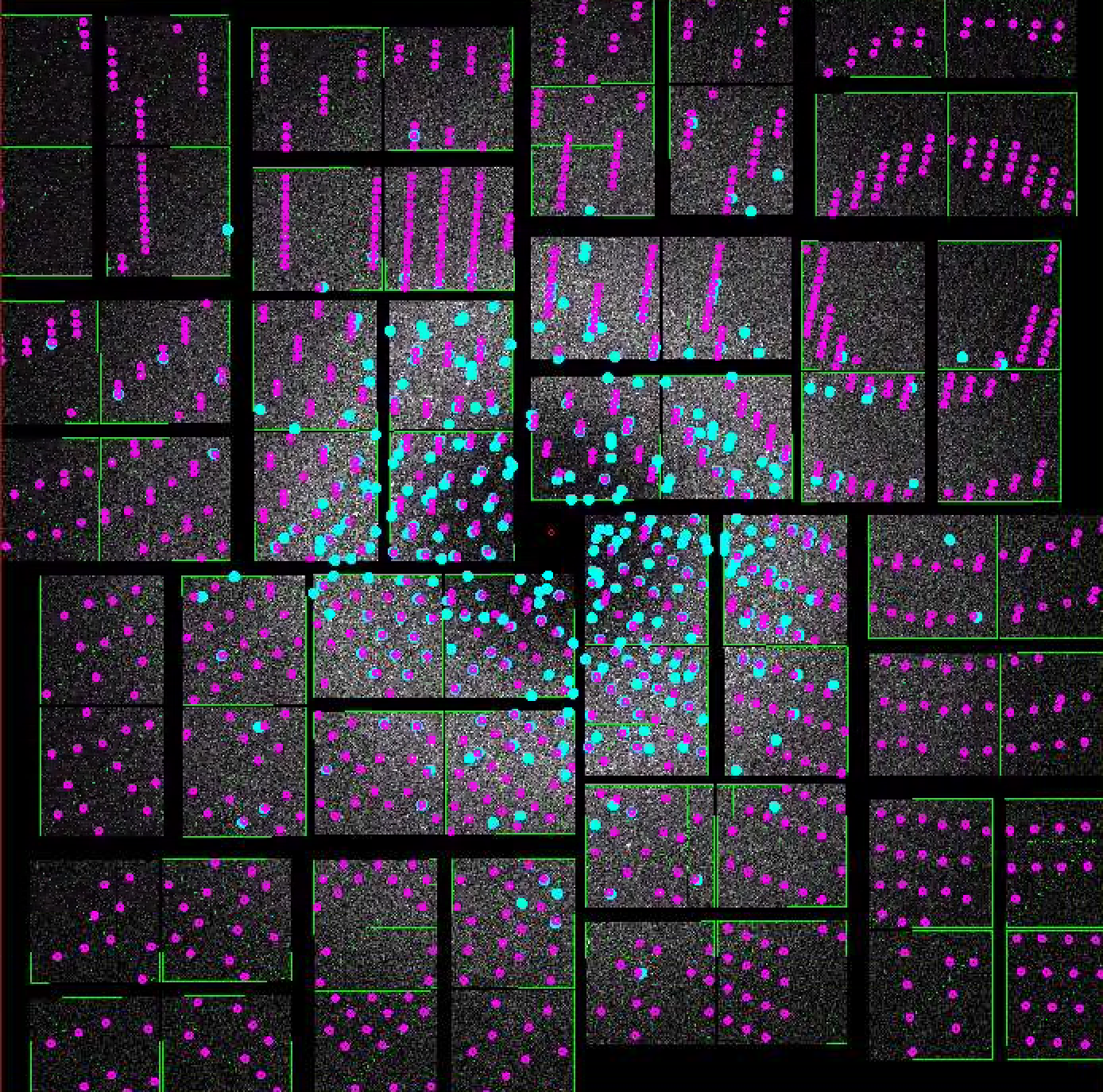}
    \caption{Diffraction pattern from a streptavidin crystal. Peaks found by the peak finding algorithm (cyan) and integrated intensity HKL positions after indexing (magenta) are shown.}
    \label{fig:roibin:diffraction}
\end{figure}

Data processing for serial crystallography consists of four steps. First, Bragg spots in the diffraction pattern (formed by the crystal lattice) are identified by using a peak-finding algorithm. Next, an indexing step determines the unitcell parameters and the orientation of each crystal based on the peak positions. Therefore, accurate measurements of both the position and intensity of Bragg spots are critical. Tuning peak-finding parameters that produce accurate peak positions in real time can be a challenge, although default parameters tend to give decent results. An example indexed diffraction pattern is shown in Figure~\ref{fig:roibin:diffraction}. The third step involves merging 2D diffraction patterns into 3D diffraction volume based on the recovered orientations. The diffraction volume represents the Fourier amplitude of the protein electron density. The fourth step involves phase retrieval where measured Fourier amplitude and recovered Fourier phases are combined to synthesize the protein electron density. In this paper we demonstrate two popular phase retrieval methods---molecular replacement (MR) and single-wavelength anomalous diffraction (SAD)---to demonstrate common cases encountered at beam lines.

%\robert{define terms for crystalography: steps of the process, what they do why important, which one are hardest to get right under lossy compression, image at the end is the quality control}

\section{Related Work}
\label{sec:relate}

Currently in serial crystallography, lossless compression is used in some contexts.
Specifically, DEFLATE-3 compression was used on the CXI files generated at LCLS, being applied event by event to the images from the detector.
These are not the long-term archival storage and often are stored in addition to the raw data files already generated at the site.
This compression technique operates by identifying patterns in the dataset and then encoding their values into a single symbol that represents the pattern.
DEFLATE-3 is so-named because it is the ``3rd level" of DEFLATE, indicating a certain window size and patience parameters used in identifying patterns.  Higher values generally indicate slower compression but higher compression ratio, while lower values generally indicate faster compression but lower compression ratios. 
During the data processing, each dataset is partitioned into multiple HDF5 chunks, each of which is handled by an MPI rank with a collective write operation. The compression is performed via the HDF5 filter mechanism \cite{thehdfgroupHDF5DataFlow2006}, followed by a parallel data writing collectively to disk using MPI-IO.
This model suffers a number of potential limitations, which can restrict throughput and scalability of compressors, as discussed in \cite{underwoodProductivePerformantGeneric2021}.

Lossy compression of light source datasets has also been studied for years. 
In 1998, Ferrer et al. \cite{Ferrer1998} investigated the impact of different lossy compressors designed based on different transforms such as wavelet transform and cosine transform, using the crystallography datasets generated by the detector of the D2AM beamline at the European Synchrotron Radiation Facility. Their studies, however, do not take advantage of the many state-of-the-art lossy compressors such as SZ and ZFP developed in recent years, and their datasets may not reflect modern detectors with greater resolution and possibly different noise patterns. 

Lossy compression for rotational crystallography \cite{jamesh} has been demonstrated to be effective by exploiting the correlations in orientations. In serial crystallography, however, each crystal orientation and fluctuation in the beam are random,  making it difficult to apply this technique.
%Lossy compression algorithms keep being developed for various  crystallograpy datasets in recent years. 

%There are two areas of related work for the use of lossy compression for lightsource data.
Hadian-Jazi et al. \cite{hadian-jaziDataReductionSerial2021} explored the use of non-hit rejection enabled by specialized peak finders.
One can discard certain events from the stream generated by the detector.
We view this as an orthogonal area of research that complements our approach and can be used in conjunction.

Another line of research  looks at lossy compressors for peak data \cite{maroneImpactLossyCompression2020, leonarskiJUNGFRAUDetectorBrighter2020} regarding light source datasets.
Leonarski et al. studied the effectiveness of the SZ compressor on the data generated by a JUNGFRAU detector\footnote{We use this detector type and dataset type for lysozyme later in the paper.}\cite{leonarskiJUNGFRAUDetectorBrighter2020}. They confirmed that while large compression ratios are possible (up to 168$\times$) with SZ 2.1.7,\footnote{\cite{leonarskiJUNGFRAUDetectorBrighter2020} reported compression ratios relative to float32, whereas we base compression ratios on the raw uint16 data from the detectors and have converted the results in this paragraph to be consistent with our reporting.} this can result in poor quality, especially in regions with small local intensity fluctuations. Thus, their solution can  use only relatively small error bounds to ensure the reconstructed data quality, which inevitably results in very low compression ratios (up to $10\times$). 
%which they characterize as having modest impact, but leave detailed quality analysis to future work.
By comparison, our solution attempts to resolve some of the quality challenges faced by this work at high compression ratios and perform more detailed quality analysis. 
%\robert{Chuck, the paper by leonarski says the JUNGFRAU natively uses a uint21, is this true of our data?, does only cspad use 16 bits?, if so should we just use 32 bit compression everywhere for consistency in the tables and figures and comment on this in the text?, RESOLVED: there is manipulation done internally}

Marone et al. leveraged JPEG2000 and JPEGXR for the general compression of visual images and applied it to a different form of light source data X-ray tomography \cite{maroneImpactLossyCompression2020}. Visual inspection of these shows that the corresponding images are substantially less noisy than the types  of data seen in crystallography, indicating  important differences between different types of light source data that may impact the suitability and effectiveness of different compression methods.
Other works such as \cite{Roy2021}, attempt to compress this type of data using deep neural networks.

Huang et al. \cite{Huang2021} considered yet another class of light source data: x-ray ptychographic data.
They applied a rounding scheme to make the data easier to compress.
Again this data is still not as noisy as the crystallography data we study in this paper. By comparison, the crystallography data is very hard to compress. Our solutions can also be applied to the ptychographic data, an area that will be investigated in our future work.

%There have been some works that attempt to look at lossy compression of crystallography data.
%This work has looked at binning + rounding + lossless compression (without region of interest preservation).
%To the best of our knowledge this work did not consider the reconstruction of electron densities (see ~\ref{fig:roibin:electron_origional}).
%Reconstruction of electron densities is a time consuming and partially manual process, but a key element of processing and understanding results from crystallography.
%We find the levels ($\epsilon_{abs}=100$, bin=$3\times3$) and methods of compression proposed in this work would fail to reconstruct the electron densities resulting in a worse than what we show in Figure~\ref{fig:roibin:electron_greedy} while producing vastly inferior compression ratios albeit more quickly.
%Our approach introduces three key concepts for achieving high quality results: 1) region of interest preservation near dynamically located peak values 2) use of more sophisticated error bounded lossy compressors 3) systematic search for optimal compression at desired quality.
%We additionally parallelize and scale the code to run efficently on HPC.

\section{Overview of the Design of \projectName{}}\label{sec:design}

\begin{figure}
    \centering
    \includegraphics[width=\columnwidth]{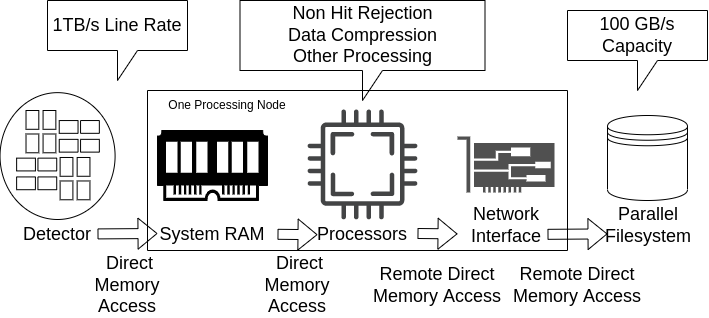}
    \caption{System dataflow overview for future LCLS-II-HE.}
    \label{fig:roibin:systemoverview}
\end{figure}

\projectName{} was designed for use in upcoming high-data-rate facilities such as LCLS-II-HE depicted in Figure~\ref{fig:roibin:systemoverview}.
In this system, each portion of the detector has the ability to write directly to RAM on a preprocessing node.
The processor can then access the data from the system RAM for processing.
The proposed design of LCLS-II-HE will operate at 1 MHz beam rate (potentially collecting up to hundreds of thousands of diffraction patterns limited by detector readout) with the incoming data stream arriving at a rate of 1 TB/s.

Once data is loaded from the detector for processing, the processor will handle multiple tasks, including detector calibration (pedestal, common-mode, and gain correction), peak-finding, NHR, data compression, and other forms of preprocessing.
First, each panel is calibrated to account for known noises in the detector.
Such a process involves converting the raw 16-bit unsigned integer data into 32-bit IEEE floating-point data, effectively doubling the input data rate.
This detail is important because when compression ratios are reported later in this work, they are reported relative to this 16-bit unsigned integer data rather than the 32-bit floating-point data that is passed to the compressor.

The peak-finding process then locates peaks within a detector panel.
This algorithm is performed separately and is needed for NHR preformed earlier in the pipeline, and optimizing peak-finding further is out of scope now. For our current work we use the well-established peak-finding algorithm \texttt{peak\_finder\_v3}.
If the peak-finding algorithm detects fewer than some number of peaks (set to 10 in our studies), the data from this event is discarded---NHR.
Based on our observations as well as some published work on peak finding \cite{hadian-jaziDataReductionSerial2021}, this process would typically discard 20 to 80\% of the incoming events, leaving a stream of $\approx$  500 GB/s that needs to be further reduced. %\robert{Cite Needed for 20 to 80\% reduction}.

\begin{figure}
    \centering
    \includegraphics[width=\columnwidth]{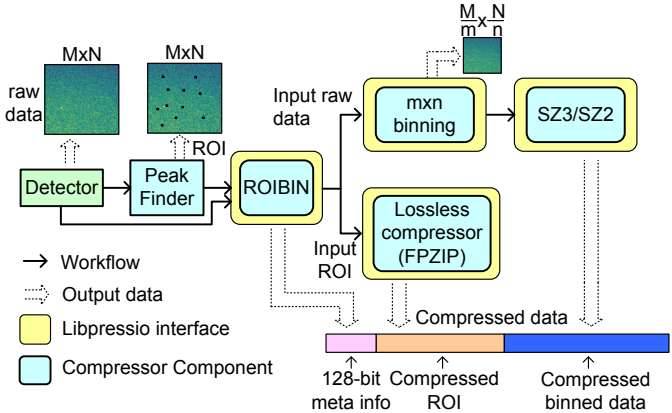}
    \caption{Compression methodology overview}.
    \label{fig:roibin:softwareoverview}
\end{figure}

This stream is  then compressed with \projectName{}, as illustrated in Figure~\ref{fig:roibin:softwareoverview}.
We first bisect the incoming data stream into the region of interest and the background.
For each peak located during peak finding, we save a rectangular region around the peak.
We choose the region of interest size to be twice that of the peak size + 1 (17x17).  We choose this size to prevent possible overlaps of bounding boxes for nondetected peaks from being treated as background information. We leave further refinement of this to future work.

We then apply lossless compression to these regions of interest. In our case we found that the lossless compressor fpzip---a specialized compressor for floating-point 
values---achieves the greatest compression ratio for this data.
We note that the maximum number of peaks in each event is  small $\lesssim 2,048$ relative to the size of the background of the detector, and in practice a much smaller number of peaks is found in each event.
This means that relative to the detector the region of interest is at most $5\%$ and often less than $1\%$ of the data to be compressed for each event and has a modest impact (at most 3.4\%---10 ms---in the worst case, 2.2\% in the median case) on the overall compression ratio. Thus does not benefit as much from further optimization per Amdahl's law.
Therefore, and as explained later in Section~\ref{sec:performanceopt}, threading the compression of the regions of interest once collected provides little benefit relative to the cost to spawn the threads in many---but not all---cases.
Therefore we switch dynamically between a serial and parallel implementation of the region of interest preservation code depending on the number of events.

After lossless preservation of the regions of interest around the peaks, we apply binning followed by lossy compression. This background information is still important for weak Bragg peak integration, namely, peaks that were too weak to be detected by the peak-finding algorithm (see integration areas (magenta) in Figure~\ref{fig:roibin:diffraction}).
This background information can be  noisy, resulting in poor compression performance for lossless and lossy compressors alike.
We  found, however, that binning reduces the volume of the data substantially while smoothing the underlying data enabling lossy compressors to achieve even higher compression ratios without impacting the scientific quality of the data.
We discuss the quality assessment to support our claims in Section~\ref{sec:qualityassesment} as well as a performance assessment of different configurations we tested in Section~\ref{sec:performanceassesment}.

We implemented and optimized a parallel algorithm for binning the background data that is then lossily compressed and the region of interest extraction that is losslessly compressed. 
Since many parts of the algorithm can have performance parameters, we use an autotuning approach to determine how to allocate these sparse resources.
We discuss the optimizations and allocations used here in Section~\ref{sec:performanceopt}
The underlying lossy compressor can also implement thread parallelism.

Our implementation makes use of LibPressio---a generic interface for lossless and lossy compression of dense tensors---to compose the phases of our implementation \cite{underwoodProductivePerformantGeneric2021}.
This allows us to quickly experiment with many possible compressor configurations for each stage of our process and consider alternative approaches without changes to the underlying code.
This also makes our implementation robust to improvements in compressor design, such as the development of application specific integrated circuits, field programmable gate arrays (FPGAs), and GPU-based compressors that are just now gaining adoption for compression.
However, because of the current system depositing the data from the detector in System RAM rather than device memory using a system such as GPUDirect,  the overall bandwidth of these compressors is limited by host-to-device and device-to-host memory transfers \cite{nvidiacorperationGPUDirect2015}.
Were this to change, our implementation would be prepared to take advantage of this change.

\subsection{Performance Optimizations}\label{sec:performanceopt}

When compressing more than one event at a time, there are some benefits in parallel execution of the extraction and restoration of the regions of interest from the background.
Since the number of regions of interest and their size are known at the time of compression, we allocate the number of peaks $\times$ the size of the region of interest.
After that we parallelize copying of each peak and its surrounding with multiple threads.
If a region of interest extends past the edge of a panel, the remaining regions are replaced with a constant fill value.
The copying of single regions of interest is also optimized.
We dispatch to specialized dimension aware copy routines that are customized for 3D and 4D data layouts.\footnote{i.e., if the data is divided into a panel-aware format (events $\times$ panel $\times$ row $\times$ column) or a merged 3d format (events  $\times$ row $\times$ column)}
These optimized versions are customized to remove unnecessary arithmetic operations when higher-dimensional indexes are known to be one and conditional jumps that would never or always will happen and were not performed by the optimizing compiler.
The code also inspects the size of the loop indexes and uses smaller iteration types (i.e. \texttt{uint16\_t} instead of \texttt{size\_t}) when they will not cause overflow because smaller index types increment in fewer cycles.
Collectively, these optimizations for datasets containing a sufficient number of peaks to warrant parallel execution yield a speedup of $2.4\times$ for extraction and $1.6$ for restoration when using four threads over serial execution.

Likewise for the binning operations we are able to make a number of optimizations.
Based on the bin size and dataset size, we can preallocate sufficient memory for the binned version of the data and divide the calculation of particular bins to individual threads.
Then, like the region of interest extraction and restoration, we dispatch to specialized routines depending on the dimension and iteration index sizes to reduce the volume of arithmetic operations and unnecessary conditional jumps.
Unlike the region of interest optimization, binning is always applied to the whole chunk and is generally parallelized for each invocation of the compressor.
These optimizations yield a speedup of $4.7\times$ and $2.2\times$ on the restore operation over serial execution.

Next, we leverage LibPressio's ability to do zero-allocation compression and decompression where possible and instead use preallocated compression and decompression buffers.
This approach reduces the latency of individual compression tasks for compressors that support it, such as ZFP and soon cuSZ, where it will be even more important due to allocation overheads on GPUs.
The benefit  can be seen partially in Table~\ref{table:roibin:performance:both} in the \texttt{roibin\_zfp} and \texttt{zfp} configuration and its particularly high compression bandwidth given that the ZFP portion of the compression is not executed in parallel in this example and is doing meaningful compression relative to \texttt{blosc} with a compression ratio of 11.44.

Many tunable performance parameters  do not affect the compression ratio or values in the reconstructed data stream but can impact the runtime of the algorithm (i.e., the number of threads).
Because we export the various elements of our process as LibPressio compressor modules, we can utilize the capabilities of LibPressio's OptZConfig to automatically determine reasonable settings for these parameters.
OptZConfig uses black-box nonlinear optimization techniques to drive a control loop to determine the optimal setting.
More on OptZConfig can be found in \cite{underwoodOptZConfigEfficientParallel2022}.

\subsection{Dynamic Selection of Regions of Interest}\label{sec:regionofinterestselection}

Regions of interest changes per image because of the random orientation of the crystal. We employ a peak-finding algorithm called \texttt{peak\_finder\_v3} \cite{peakfinderv3}. The algorithm starts by identifying all local intensity maxima within a sliding window of size 7x7 pixels (set proportional to the typical size of a Bragg peak). Local maxima below 300 ADUs are rejected. All neighboring pixels above 0 ADUs are considered part of the peak. Total peak intensity must be greater than 600 ADUs, and signal-to-noise ratio must be greater than 10. The number of neighboring pixels must be between 2 and 30 pixels. All these parameters are configurable by the user.

Peak finding was performed by using MPI in psocake \cite{Damiani:zw5004,Thayer2017}. For a 2-megapixel cspad detector, it took an average of 65 ms for detector correction and 100 ms for peak finding per image on Intel Xeon CPU E5-2620 v3 @ 2.40GHz. For a 4-megapixel jungfrau detector, it took on average 480  ms and 200 ms for detector correction and peak finding, respectively. Because of the complexities in automatic gain switching, the jungfrau takes longer to calibrate compared with a single gain cspad.  Algorithmically,  however, peak finding scales linearly with the number of pixels.
%\robert{full image with peaks annotated, lysozyme background histogram+peak histogram}

Serial crystallography is not the only discipline that can take advantage of region of interest preservation.
The results of serial crystallography instruments inform the science of other domains including structural biology and materials science.
We also expect that this technique will translate to related techniques such as ptychography.
Additionally, other applications such as the preservation of local or global extrema \cite{liangFeaturePreserving2D3D2020} or the preservation of key features of a scientific dataset (e.g.,   tropical cyclones in a climate dataset \cite{bakerEvaluatingImageQuality2019})  are important aspects that affect other domains of scientific work.
Such aspects  include the relative proportion of regions of interest and tools to automatically identify them with a low miss rate (false negative/positive).
Thus our implementation also supports 2D and 1D inputs and arbitrary hyper-rectangles for the region of interest.
Users can provide region of interest information as an array of coordinates, allowing this technique to be used outside of this domain.
However, validating this approach in each domain is a highly fact-bound process requiring extensive feedback from domain experts. Thus we focus on serial crystallography for this work, leaving further expansion to future work.

%Timing info:
%cspad2M (32,185,388):
%calibration 65ms per image
%=15.38 images/s
%=141.32 MB/s
%peak finding 100ms per image
%=10 images/s
%=91.88 MB/s

%4,326,979 total events
%744,150 hits (17.2%) CR = 5.81 
%jungfrau4M (8,512,1024): 
%calibration 480ms per image
%=2.08 images/s
%=34.90 MB/s
%peak finding 200ms per image
%=5 images/s
%=83.89 MB/s
%248,024 total events
%77,120 hits (31.1%) CR = 3.22

\section{Experimental Setup}
\label{sec:experimentalsetup}

%\robert{which cxi file name is what substance}
% selenbiotioyl (Se-SAD)- cxic0415
% lysozym - cxic00318

\subsection{Hardware and Software}
We conduct all of our experiments on up  to 10 identical nodes with the hardware and software described in Table~\ref{tab:roibin:hardware_software}.
We choose this hardware because it was available on the cluster used for testing and is similar to hardware that what may be used in the compression processing component for LCLS-II or the updated APS.
The software generally represents the latest versions available in spack at the time of testing, built using default optimization levels with spack (generally \texttt{-O2})with an extra repository added for compressors \cite{underwoodRobertu94SpackPackages2020}.
However, we used the system compiler and system MPI packages because they are optimized for the particular hardware.

\begin{table}[ht]
    \centering
    \caption{Hardware and software configuration}
    \label{tab:roibin:hardware_software}
    \begin{tabular}{lr|lr}
    \toprule
       Component & Version & Component & Version \\
    \midrule
       CPU & Intel Xeon 6148G (40 cores) & sz2 & 2.1.12.2 \\
       GPU & 2 Nvidia v100 & sz3 & 3.1.3.1 \\
       Memory & 372 GB & zfp & 0.5.5 \\
       Network & 2 Mellanox MT27710 (HDR) & mgard & 1.0.0 \\
       FileSystem & BeeGFS 7.2.3 (24 targets) & blosc & 1.21.0 \\
       Compiler & GCC 8.4.1 & fpzip & 1.3.0 \\
       OS & CentOS 8.2.2004 & bzip2 & 1.0.6 \\
       MPI & OpenMPI 4.0.5 & HDF5 & 1.12.1 \\
       LibPressio & 0.83.2 & OptZConfig & 0.13.1 \\
    \bottomrule
    \end{tabular}
\end{table}

%\robert{TODO(chuck) can you please fill dataset details in here?}
\subsection{Datasets}
We conducted quality assessments on two independent datasets. The first dataset is selenobiotinyl-streptavidin on a cspad detector with a total of 4,326,979 images where 744,150 were contained hits (17.2\%). The compression ratio for only non-hit rejection is the inverse of the hit rate, CR = 5.81. The second dataset is lysozyme on a jungfrau4M detector with a total of 248,024 events where 77,120 were hits (31.1\% and non-hit rejection only CR = 3.22). 
These detectors have different resolutions (4096, 1024) and (1480, 1552) respectively.

Additionally, these two datasets  represent two important use cases: Se-SAD phasing technique is used for reconstructing streptavidin, which means that a weak anomalous signal from selenium atoms has to be detectable after decompression, making this an ideal dataset to test sensitivity of our lossy algorithm. The lysozyme dataset uses molecular replacement, which is the most common phasing technique used at beamlines.

For the experiments in quality assessment (Section ~\ref{sec:qualityassesment}), we use entire runs of the experiments because using the entire dataset is essential to the assessment of the quality of the approach.
However, for experiments in the performance assessment (Section~\ref{sec:performanceassesment}), we use only a subset of a few hundred images to quickly assess a variety of methods.

The datasets considered are limited by the fact that end-to-end evaluations consume enormous amounts of resources for time and domain expert labor. The raw data for Se-SAD alone was 20TB and multiple passes over the data are required during peak finding, parameter grid-search for ROIBIN-SZ compression and decompression, indexing of decompressed data, merging into 3D diffraction volume, and phase retrieval; moving data between facilities and processing them in full for the quality analysis takes multiple days using many nodes. More work is needed to facilitate automated processing of this data.

%\robert{TODO(chuck)why are these representative?}

%lysozme is common
%SeSAD we don't start with a know model, and arrive at correct (hard)

\subsection{Compressor Configurations}

To highlight the benefits of our approach, we consider a range of different compressor configurations to represent a wide array of possible compression approaches.
We choose three leading lossless and three leading lossy compressors for the baseline of our comparison.
We additionally consider six variants on our approach that measure the performance of just binning and region of interest selection, as well as different lossy compressors for compressing the background.

We also considered several methods from related work.
NHR is a largely orthogonal topic that can be used in conjunction with our approach.
JPEG-style compression suffers from multiple challenges on this data: the multidimensional nature of the data, the range of values, and the introduction of artifacts.
The problems faced by JPEG-style compressors are largely resolved by using ZFP, which also primarily uses a near-orthogonal transform like JPEG.  ZFP is a leading lossy compressor for scientific data that we do consider in our work.
And while they are fast, our preliminary work showed that techniques such as rounding fail to reproduce the electron densities at suitable compression ratios.
SZ compression by itself---without region of interest protection---either failed to achieve a high compression ratio or failed to reproduce the electron densities.
We also found techniques that required transferring the data to the GPU for deep-learning-based inference such as TEZIP would be far too slow after accounting for just moving the data to the GPU without additional processing.

Note that we only consider the CPU versions of these compressors.
The current design of the detectors as described above starts with the data in memory that is local to the CPU, not the GPU.
This requires that the use of any GPU-based compressor begin and end with transfers to and from the CPU, which ends up being the bottleneck in these uses even though GPU-based compressors can get much higher throughput than do CPU ones if the data is resident on the GPU.
If a future revision of the design is made, our approach can easily be adapted to the GPU using the GPU compressors within LibPressio.
We present a subset of these results.

\subsubsection{Lossless Compressors}
First we describe the lossless approaches:
bzip2-6 and bzip2-9 use the bzip2 compressor with \texttt{block\_size\_10k} equal to 6 (the default), and 9 (the most aggressive). This is the most similar to what is currently used by the DEFLATE-based algorithm with compression level 3 (hereafter level; three levels are faster than the default).
Blosc-3, blosc-6, and blosc-9 use the c-blosc v1 API to invoke the Zstandard lossless compresssor with compression level 3, 6, and 9, respectively.
Z-standard represents the state of the art of general-purpose lossless compressors; c-blosc provides an ease-of-use adaptor above this that enables high-throughput compression by managing threading and other compression improvements such as shuffling.
Level 6 is the default, and level 3 represents faster and level 9 more aggressive lossless compression.
FPZip is a lossless  compressor specialized for floating-point data.
Traditional lossless compressors tend not to compress floating-point data well \cite{lindstromFastEfficientCompression2006}.
The reason  is that near values in the IEEE floating-point representation have vastly different in-memory representations.
FPZip exploits this behavior using a prediction-based approach and different encoding method that compresses this data more effectively.
Here we use only the lossless mode of FPZip because its lossy mode is generally superseded by ZFP,  one of the lossy compressors we include in our analysis.

We do not need to conduct a quality analysis for the lossless compressors---the data is not modified, and thus its results would not differ from those of the original dataset.  However, our results (Table~
\ref{table:roibin:performance:both}) show that it does not provide enough compression ratio for user requirements.

\subsubsection{Existing Lossy Compressors}
Next we consider the leading lossy compressors for scientific data.
SZ is the SZ2 series compressor that uses a combination of the Lorenzo predictor and a blockwise regression predictor with linear quantization to achieve high-quality lossy compression.
SZ3 is the evolution of the SZ2 series compressor that introduces a tricubic interpolation predictor with a larger block size.
It generally achieves greater compression ratios than SZ2, or if not greater compression ratios  and better quality at the same compression ratio.
MGARD takes a different approach, leveraging a multigrid method to determine ``nodal'' coefficients that represent the values of the data in a particular subdomain of the data within a certain tolerance; these coefficients are then compressed by using a lossless compressor \cite{ainsworthMultilevelTechniquesCompression2018}.
ZFP takes yet another approach \cite{lindstromFixedRateCompressedFloatingPoint2014}.  It partitions the data into $4^n$ blocks where $n$ is the number of dimensions of the input data.  Each block is converted to a  fixed point, a near orthogonal transform is applied, and then a specialized encoding is applied.
The near-orthogonal transform operates by converting each block from an array of values to an array of a sum of patterns that tends to be more compressible in smooth data.
For each of these compressors, we use a value range relative error $\epsilon$ bound of $1 \times 10^{-3}$, which was chosen as a conservative starting point.
The value range relative error bound is defined as follows: For each element of data $d_i$ the corresponding decompressed value $\tilde{d_i}$ will be such that $|d_i - \tilde{d_i}| < \epsilon \left(\max_{i \in D}d_i - \min_{i \in D}d_i\right)$.

\subsubsection{Our Approach}
We also consider a number of variants based on our design identified in Table~\ref{table:roibin:performance:both}  with the prefix \texttt{roibin\_}.
In these configurations we replace the background compressor with the compressor listed after the prefix.
For bzip and blosc, we use the same configurations described above.
For SZ, SZ3, ZFP, and MGARD we instead use an absolute error bound of $90$ to be consistent with our results that captured as much of the signal above the noise floor of these detectors in Section~\ref{sec:qualityassesment}.
The absolute error bound is defined as follows: For each element of data $d_i$ the corresponding decompressed value $\tilde{d_i}$ will be such that $|d_i - \tilde{d_i}| < \epsilon$. Note that since we apply binning prior to compression, the error bounds are  guaranteed only with respect to the binned data.

\subsubsection{Choice of Chunk Size}
We consider several possible event chunk sizes between 1 event and 64 consecutive events per invocation of the compressor.
The former more closely maps to the current design where each event is stored in its own HDF5 chunk.
The latter attempts to provide more information to the lossy compressor to attempt to exploit correlation in the background information between events.
This might lead to the intuition that the chunk size should be some even division of the total number of events. Two constraints, however,  lead away from this conclusion.
First, these datasets and auxiliary peak-related metadata are sufficiently large ($\approx 6TB$ for cxic00318\_0123), and the entire dataset is unlikely to fit into main memory at the same time even if spanning a small number of analysis nodes, especially after accounting for other working memory for the analysis tasks and compression operations.

Second, during the analysis phase, users often access individual events interactively to better understand the structures of the substances being considered.
This implies that data decompression needs to be quick in order to not impair interactive use, and large chunk sizes contraindicate fast access to individual events.
We choose several event sizes to explore the benefits of larger chunk size while balancing available memory and interactive use with compression sizes.

\subsubsection{Autotuning Configuration}

For our autotuning, we allowed each aspect of the system to consider between 1 and 8 threads of execution for each aspect of execution and between 1 and 40 for the number of tasks.
We allowed the tuner to use up to $\sqrt{n}$ iterations, where $n$ is the number of non-task parallelization points for the configuration, and to not use early termination.
This approach gives OptZConfig the opportunity to search a substantial-enough portion of the search space without testing every configuration.
We then ask the tuner to find a minimum compression time for each configuration.

Since the tuning system is executed multiple times---once for each separate task---we have several possible thread allocations for each rank.  We choose the thread allocation that was most common for the configuration to be the thread allocation for the whole.  This allows for some level of run-to-run variability that is not well accounted for in~\cite{underwoodOptZConfigEfficientParallel2022}, which was defined for mostly consistent quality optimization rather than performance optimization.

\section{Quality Assessment of \projectName{}}\label{sec:qualityassesment}

%Discuss the search for optimal quality assessment parameters
Three key parameters  affect compression ratio: binning, tolerance, and dimension.
Binning is the dimensions of window of data values which are averaged together in the binned representation, i.e. $(3\times3)$ combines 9 values and $(1\times1)$ skips binning.
Tolerance is the maximum absolute point-wise error allowed during compression of the \textit{binned} values \footnote{This means that the maximum absolute point-wise error on the raw values can be much larger up to tolerance + 50\% of the value range, in practice this is mitigated by the preserving areas around Bragg spots which induce large binning errors}.
Dimension refers to the dimension in which adjacent features in the data are exploited by SZ for compression. A few pixels at the beginning and end of a feature can overlap with adjacent end and beginning feature  along the row of pixels (1D), entire panel (2D), or stack of panels (3D).

\begin{figure*}
    \centering
    \begin{subfigure}{\textwidth}
        \centering
        \includegraphics[width=.85\textwidth]{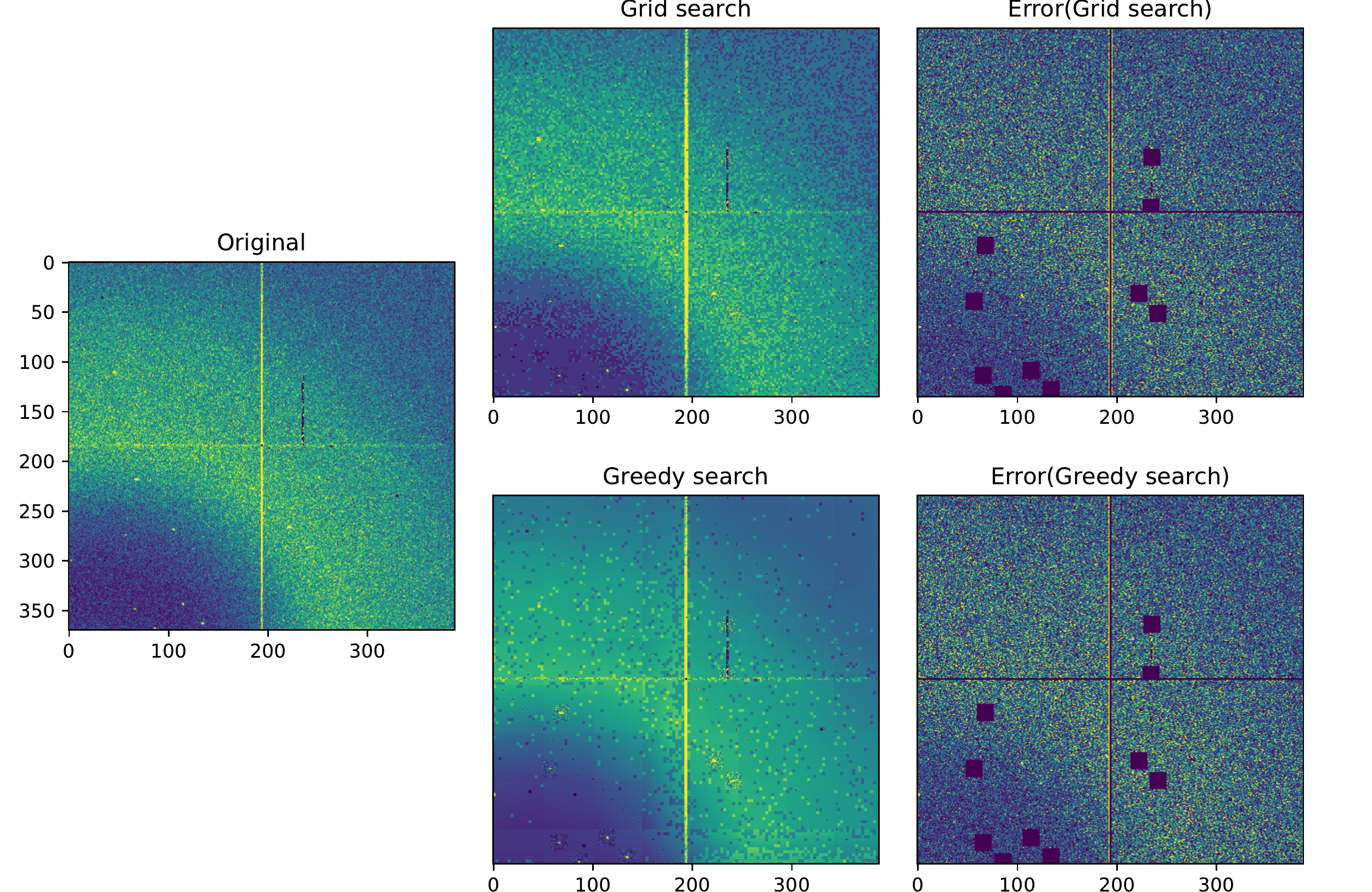} 
    \end{subfigure}
    \caption{Selenobiotinyl-streptavidin: Search for optimal compression ratio parameters in Table~\ref{tab:roibin:quality_search}. Decompressed image and error from best grid search: binning = 2, tolerance = 90, and compression dimension = 3 (top) Decompressed image and error from greedy search: binning = 3, tolerance = 90, and compression dimension = 2 (bottom).}
    \label{fig:electron_density_sadse}
\end{figure*}

\begin{figure*}
    \centering
    \begin{subfigure}{0.32\textwidth}
        \centering
        \includegraphics[width=\textwidth]{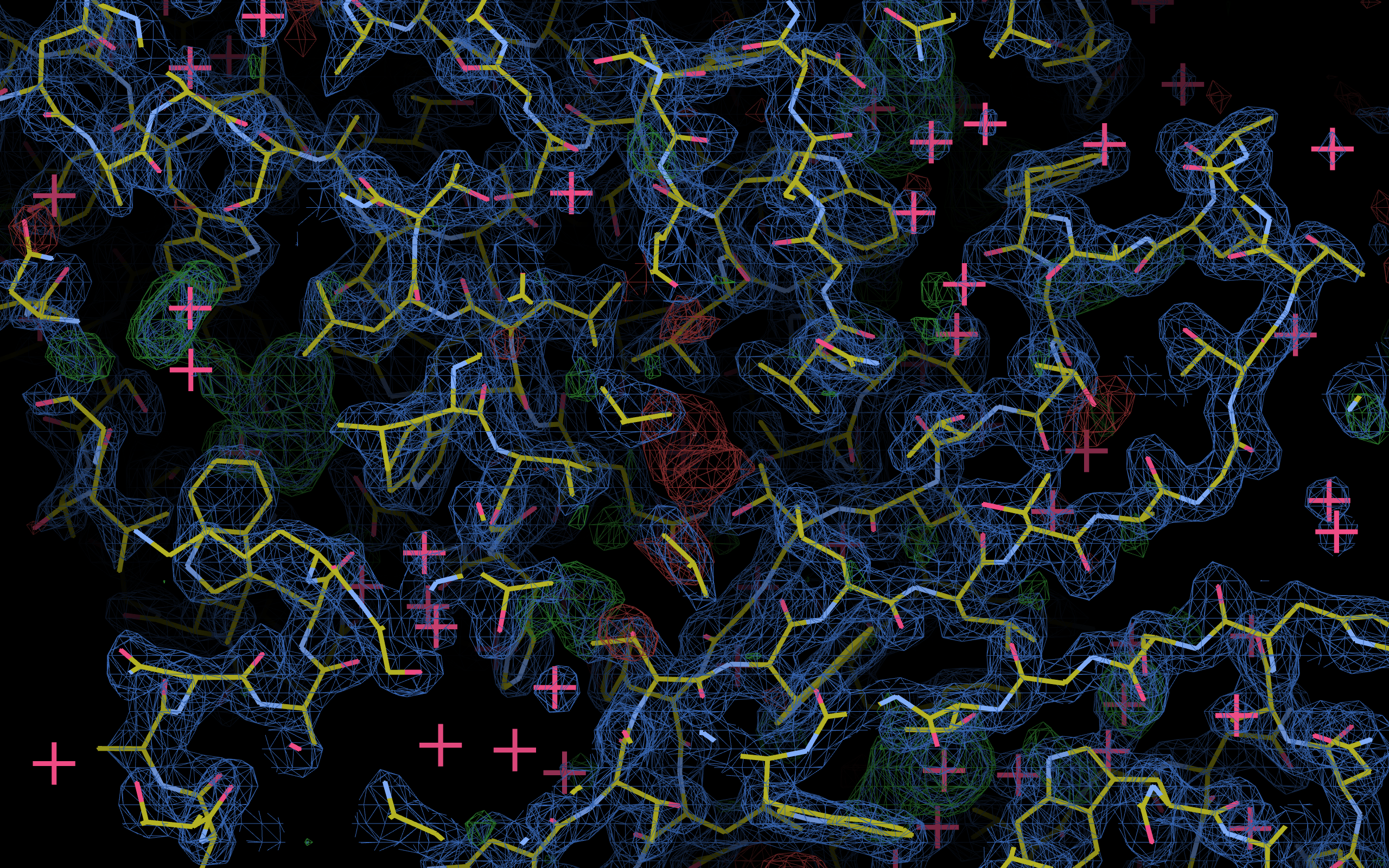}
        \label{fig:roibin:electron_grid}
        \caption{original}
    \end{subfigure}
    \begin{subfigure}{0.32\textwidth}
        \centering
        \includegraphics[width=\textwidth]{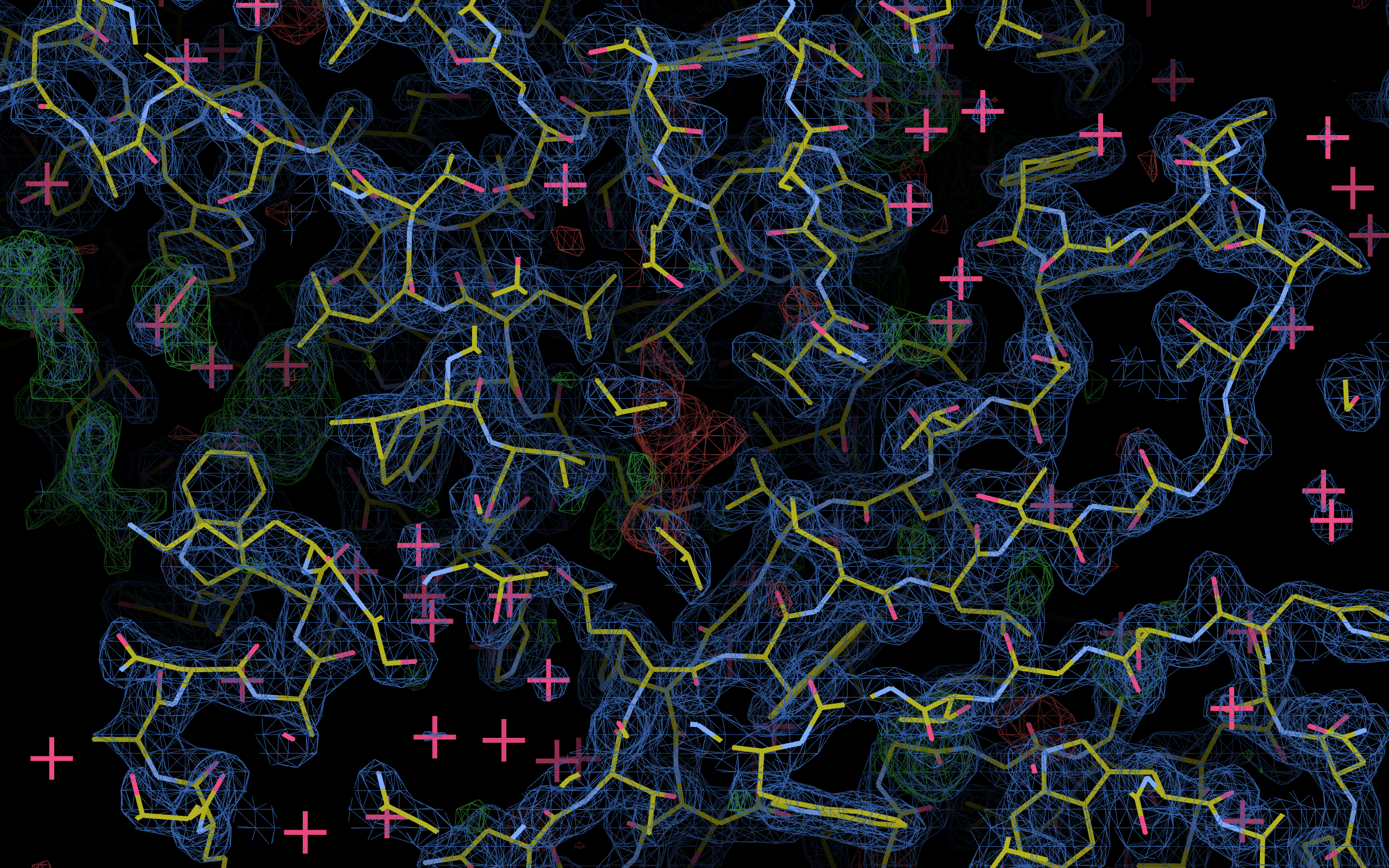}
        \label{fig:roibin:electron_origional}
        \caption{grid search}
    \end{subfigure}
    \begin{subfigure}{0.32\textwidth}
        \centering        \includegraphics[width=\textwidth]{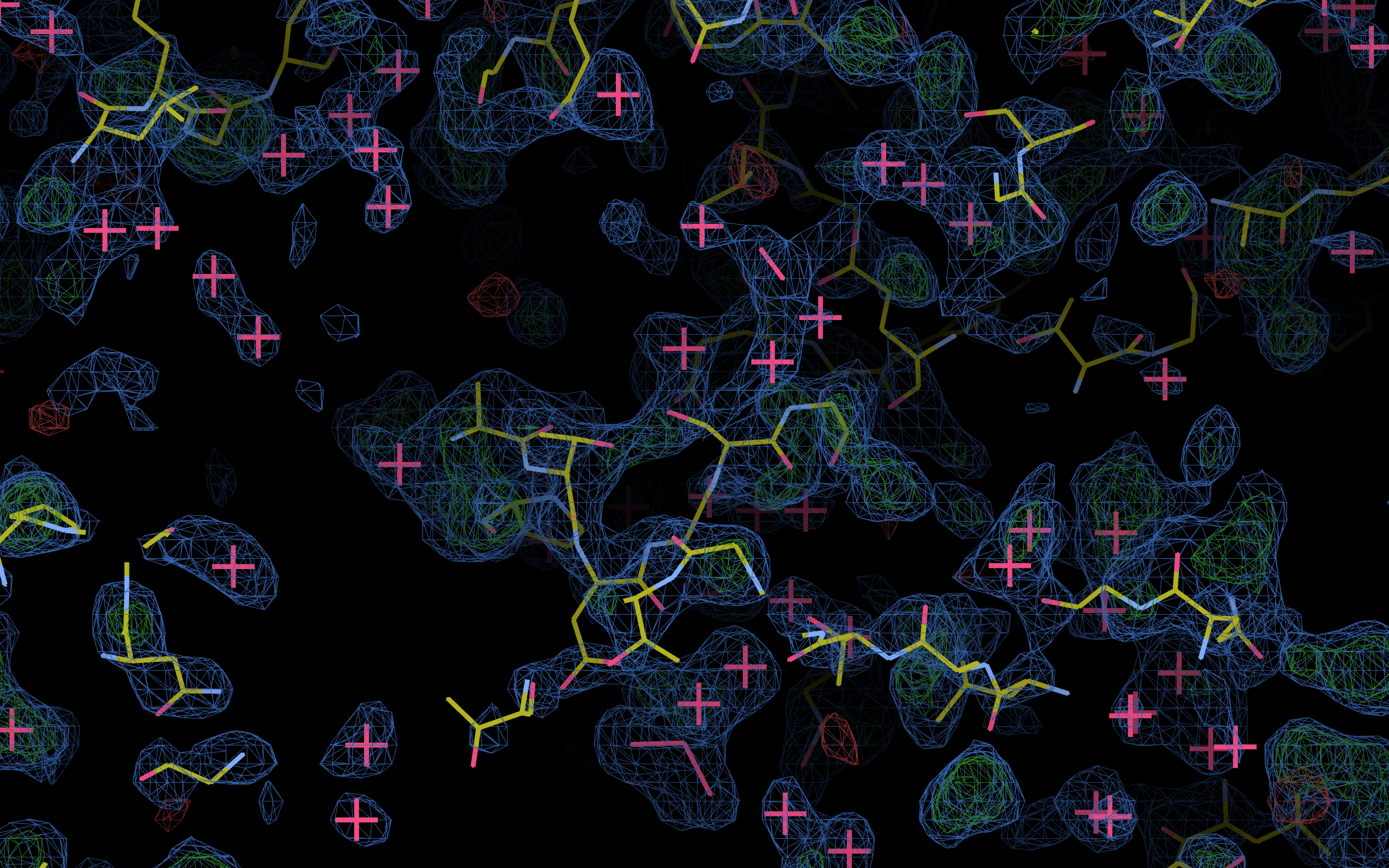}
        \label{fig:roibin:electron_greedy}
        \caption{greedy search}
    \end{subfigure}
    \caption{Selenobiotinyl-streptavidin electron density and model reconstructed using Se-SAD with original, compressed using optimal parameters from grid search, and greedy search.}
    \label{fig:roibin_cxic0415_123_electron}
\end{figure*}

Based on the typical distance between Bragg spots, we decided on 2x2 binning and also tried 1x1 and 3x3. Tolerance (absolute error bound) for the experiment is proportional to detector readout for a single photon, which was 45 analogue-to-digital units (ADUs). Search was also conducted at 10 and 90 ADUs.  We tried 1D, 2D, and 3D interpretation of the data.
A hyperparameter search for optimal compression was conducted by using a grid search as shown in Table~\ref{tab:roibin:quality_search}.
In this table the rows in the bottom section represent the parameters held constant, and the top section represents parameter that was modified -- i.e. the top left cell in the bottom section uses $1\times1$ binning, tolerance of 10, and treats the data as 1D.
The best compression was achieved for 2x2 binning, tolerance of 90 ADUs, and compression dimension of 3. A greedy search over these parameters resulted in 3x3 binning, tolerance of 90 ADUs, and compression dimension of 2. Visualization of a decompressed detector panel is shown in Figure~\ref{fig:electron_density_sadse}. The dark squares in the images showing the error are regions of interest where there are no errors. Reconstruction of the electron density is shown in Figure~\ref{fig:roibin_cxic0415_123_electron}. The quality of the electron density from the grid search is on par with the original reconstruction. However, greedy search failed to recover the correct electron density.

%\robert{Table~\ref{tab:detailed_quality_assessment} explain each quality metric, and threshold}

A subset of data was used for the grid search, resulting in a deviation from the compression ratio shown in Table~\ref{tab:detailed_quality_assessment}.
The table %~\ref{tab:detailed_quality_assessment} 
shows detailed crystallography quality metrics. The three columns original, grid, and greedy refer to data processed without compression, optimal compression parameters found via grid search in Table~\ref{tab:detailed_quality_assessment}, and compression parameters found via greedy search,  respectively. Average compression ratio is given with respect to uint16 raw data stored on disk. Note that the average compression ratio for the original dataset is 0.5 since the hits are stored in HDF5 as calibrated data in float32.
Number of hits represents the number of images that contain at least 10 Bragg peaks. Number Indexed refers to images where orientation and unitcell information could be extracted. $R_{split}$ is a measure of discrepancy between two half datasets 
\begin{equation}
R_{split} = 2^{-0.5} \frac{\sum |I_1-kI_2|}{ 0.5 \sum I_1+kI_2},
\end{equation}
where $k$ is a scaling factor and $I_1$ and $I_2$ are intensities from two half datasets.
$CC_{1/2}$ is the Pearson correlation coefficient 
\begin{equation}
CC_{1/2} = \frac{\sum^{n}_{i=1} (I_1^i-\overline{I_1})(I_2^i-\overline{I_2})}{ \sqrt{\sum^{n}_{i=1} (I_1^i-\overline{I_1})^2} \sqrt{\sum^{n}_{i=1} (I_2^i-\overline{I_2})^2}}.
\end{equation}
$CC_{ano}$ is the correlation coefficient for anomalous scattering. a measured correlation coefficient of the Bijvoet differences of acentric reflections. $R_{work}$ measures the discrepancy between the experimental observations and ideal calculated values:
\begin{equation}
R_{work} = \frac{\sum | |F_{obs}|-|F_{calc}| | }{\sum |F_{obs}|}.
\end{equation}
$R_{free}$ assesses possible overmodeling of data using the same formula as $R_{work}$ but on a small random sample of dataset aside and never included in the model refinement. Map-model CC is the correlation between model and density map. The arrows next to the metric in Table~\ref{tab:detailed_quality_assessment} indicate whether increase or decrease in the metric is preferred. As a general rule, $R_{split}$ less than 20$\%$ is a good dataset, and any value below 10$\%$ is an excellent dataset. $CC_{ano}$ less than 0.04 is a weak anomalous signal from selenium atoms, and 0.2 is a strong anomalous signal. $R_{work}$ and $R_{free}$ usually range between 0.2 and 0.6 for proteins. With optimal parameters from the grid search, the metrics show that the science is well preserved compared with the original dataset.

\begin{figure}
    \centering
    \includegraphics[width=\columnwidth]{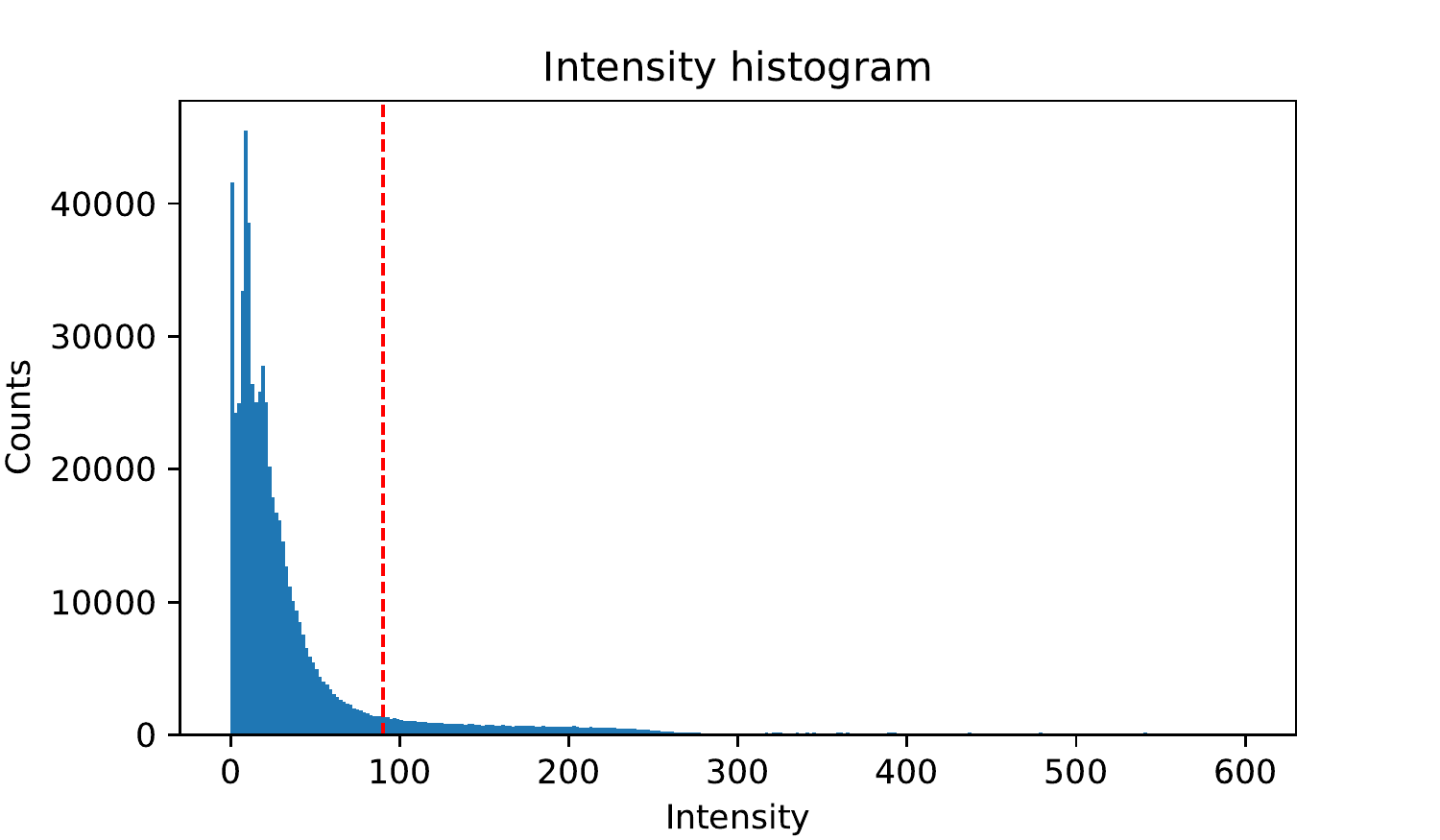}
    \caption{Intensity histogram of all pixels inside the peak window for the lysozyme dataset. Absolute error bound (red line) was set close to the elbow of the histogram. The tail of the histogram extends up to 128,517.}
    \label{fig:lysozyme_hist}
\end{figure}

\begin{figure*}
    \centering
    \begin{subfigure}{\textwidth}
        \centering
        \includegraphics[width=\textwidth]{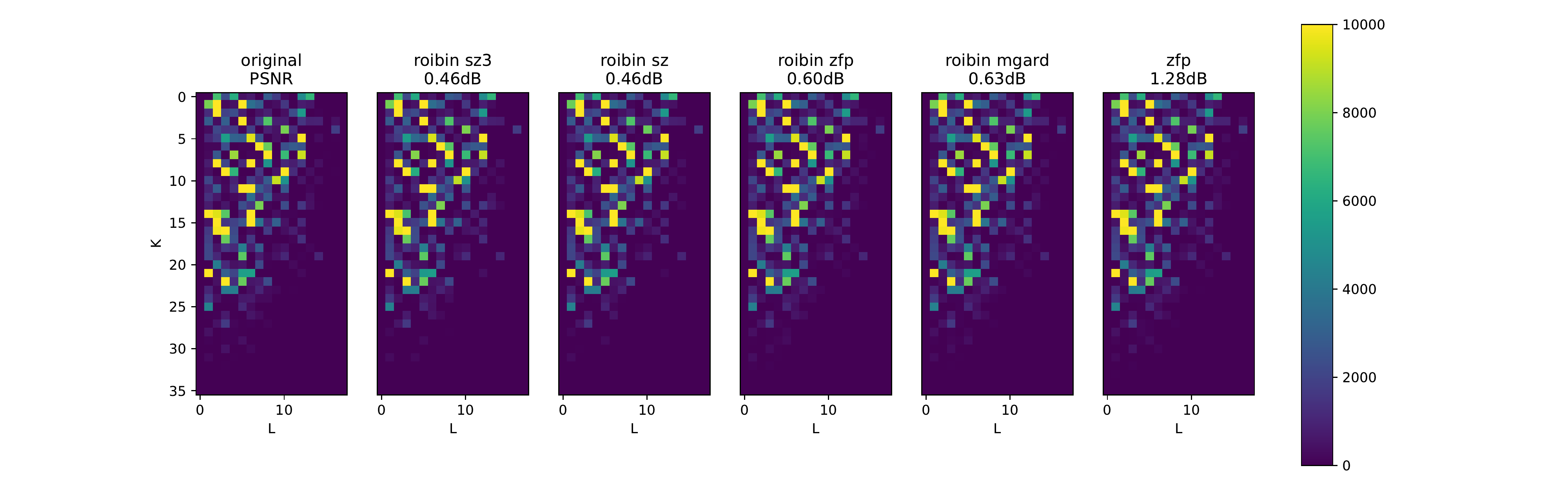} 
        \includegraphics[width=\textwidth]{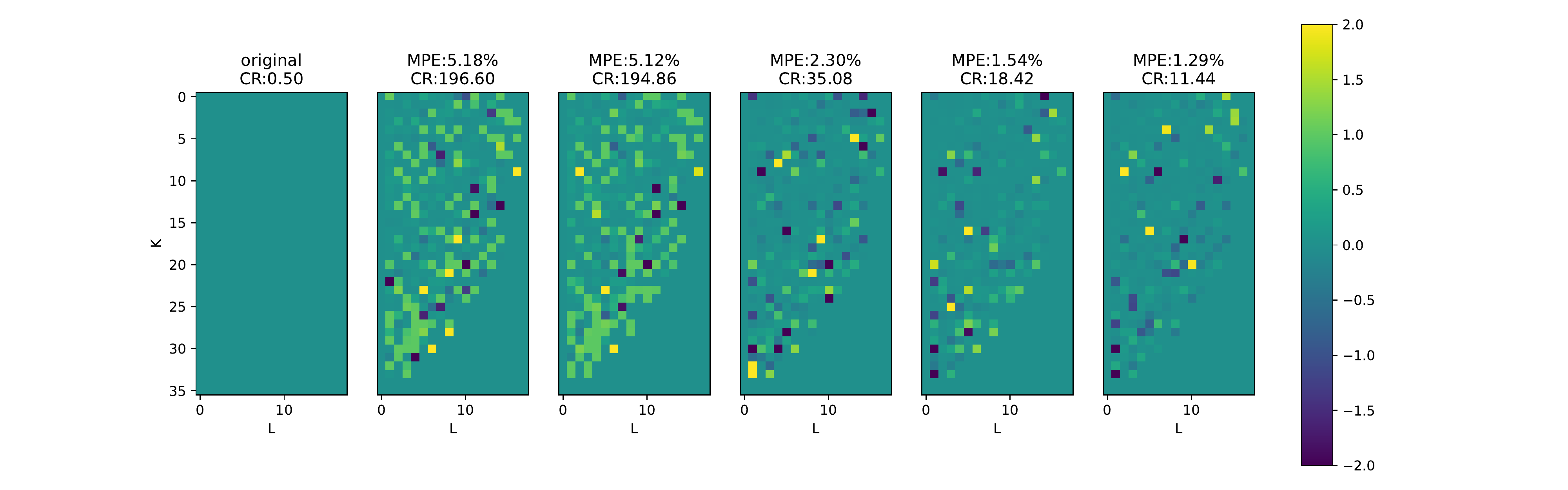}
    \end{subfigure}
    \caption{Merged intensity from lysozyme crystals for various compression methods: merged intensity HKL volume at plane H=1 and PSNR are shown using absolute error bound = 90 (top); corresponding mean percentage error MPE and compression ratio (bottom). }%\robert{chuck: could the bottom half be made as relative error instead of absolute on the gradient? $\pm 2\%$}}
    \label{fig:mergedIntensity}
\end{figure*}

\begin{figure*}[t]
    \centering
    \begin{subfigure}{0.4\textwidth}
        \centering
        \includegraphics[width=\textwidth]{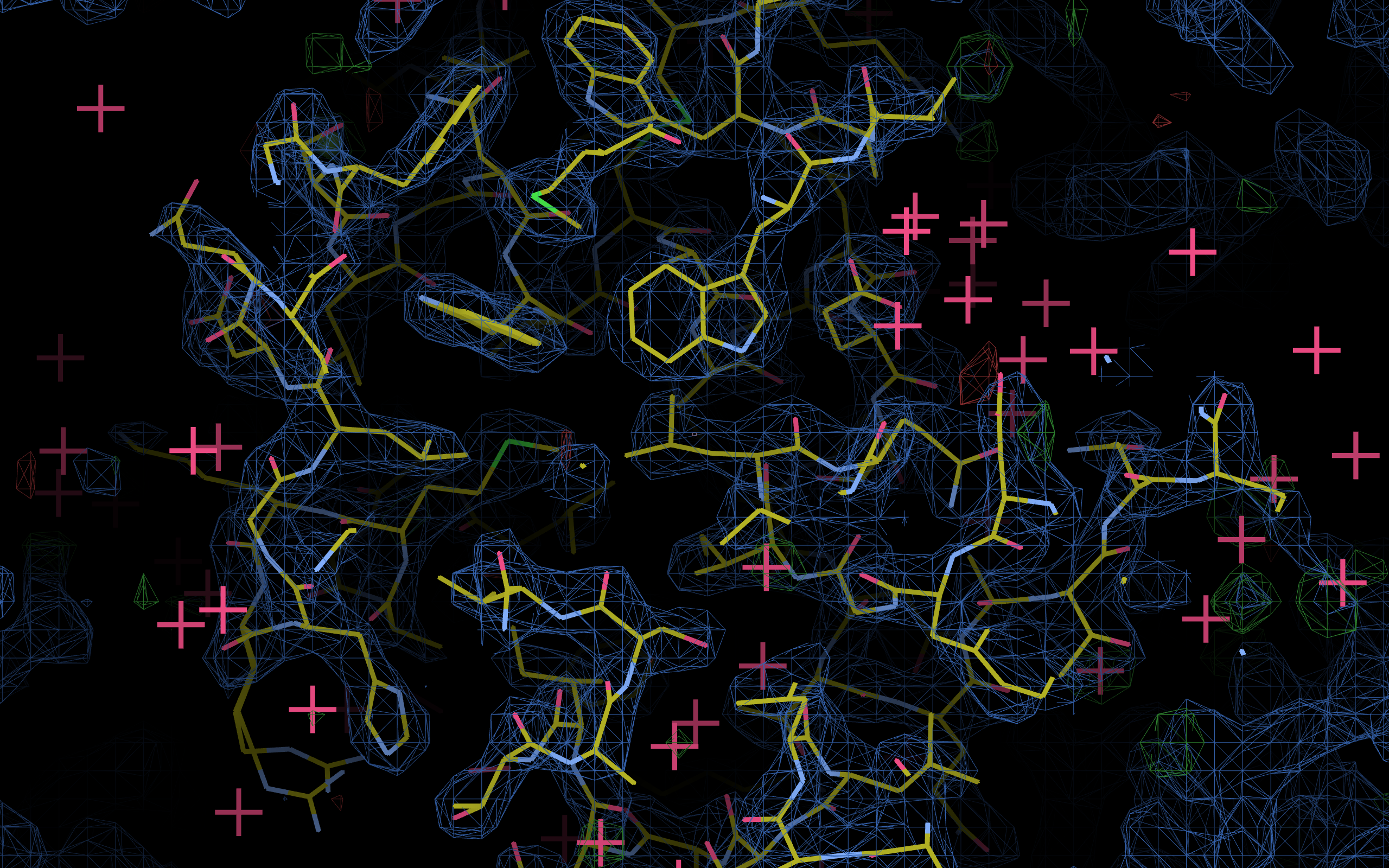}
        \label{fig:lysozyme_original}
        \caption{original}
    \end{subfigure}
    \begin{subfigure}{0.4\textwidth}
        \centering
        \includegraphics[width=\textwidth]{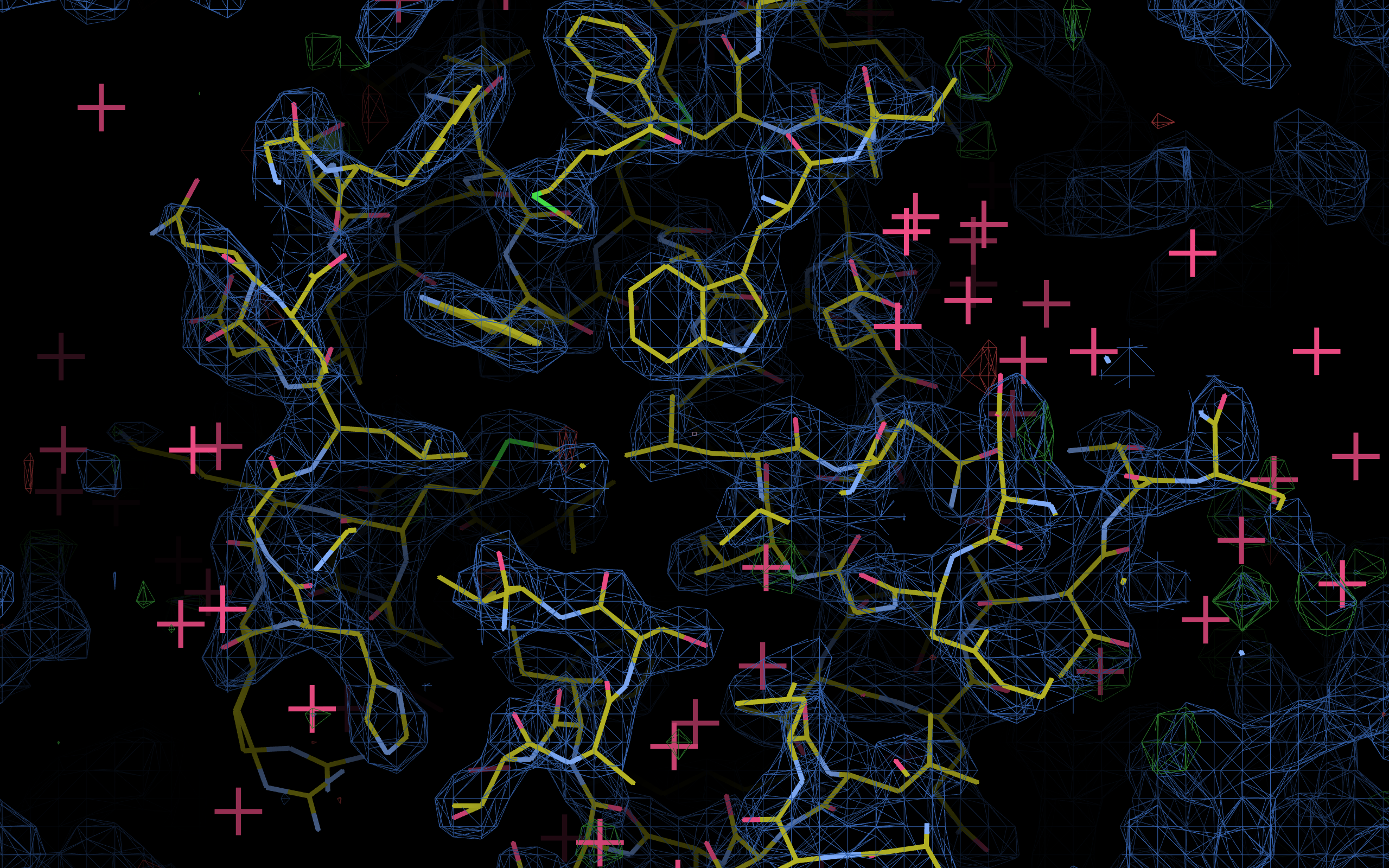}
        \label{fig:lysozyme_roibinsz}
        \caption{roibin-sz}
    \end{subfigure}
    \caption{Lysozyme electron density and model reconstructed using molecular replacement with original and roibin-sz compressed data.}
    \label{fig:lysozyme}
\end{figure*}

For the lysozyme dataset, we studied the peak signal-to-noise ratio (PSNR) and mean percentage error (MPE) of merged intensity. An absolute error bound of 90 was determined to be appropriate based on the intensity histogram (Figure~\ref{fig:lysozyme_hist}). Merging is performed on indexed diffraction patterns where we assign Miller indices to each Bragg spot (represented by three integers H,  K, and L). Each voxel of the 3D HKL volume contains averaged Bragg intensities corresponding to the same Miller indices. PSNR and a 2D plane of the HKL volume at H=1 are shown in the  top row of Figure~\ref{fig:mergedIntensity} for various compression algorithms. MPE and error are shown in the bottom row. Although \projectName{} has higher MPE than other compressors have, the compression ratio is much higher. The reconstruction from \projectName{} compared with the original reconstruction in Figure~\ref{fig:lysozyme} shows no degradation in the structure.

\begin{table}[hbt]
    \centering
    \caption{Hyperparameter quality search}
    \label{tab:roibin:quality_search}
    \begin{tabularx}{\columnwidth}{lXXX}
    \toprule
     binning (pixels) & $1\times1$ & $2\times2$ & $3\times3$ \\
     tolerance (ADUs) & 10 & 45 & 90 \\
     dims (unitless) & 1 & 2 & 3 \\
    \midrule
    tolerance 10, dim 3              & 3.11 & 12.69 & 24.99  \\
    binning, dim 3                   & 12.69 & 22.65 & 33.11 \\
    tolerance 10, binning $2\times2$ & 12.22 & 12.69 & 12.65 \\
    \bottomrule
    \end{tabularx}
\end{table}

%\begin{table}[hbt]
%    \centering
%    \caption{Hyperparameter Quality Search}
%    \label{tab:roibin:quality_search}
%    \begin{tabularx}{\columnwidth}{lX}
%    \toprule
%         binning & tolerance=10, dim=3  \\
%    \midrule
%         $1\times1$ & 3.11 \\
%         $2\times2$ & 12.69\\
%         $3\times3$ & 24.99 \\
%    \midrule
%    \midrule
%         tolerance & binning=$2\times2$ dim=3  \\
%    \midrule
%         10 &  12.69\\
%         45 &  22.65\\
%         90 &  33.11\\
%    \midrule
%    \midrule
%        dimension & binning=$2\times2$, tolerance=10 \\
%        \midrule
%        1 &  12.22\\
%        2 &  12.69\\
%        3 &  12.65\\
%     \bottomrule
%    \end{tabularx}
%\end{table}

\begin{table}[hbt]
    \centering
    \caption{Crystallography quality assessment for Se-SAD SFX Dataset. Metrics are calculated for resolution range 28\AA - 1.9\AA. Greedy search failed to preserve the integrity of the data.}
    \label{tab:detailed_quality_assessment}
    \begin{tabularx}{\columnwidth}{lXXX}
    \toprule
         Metric & Original & Grid & Greedy \\
    \midrule
         Total Compression Ratio & 2.91 & 70.65 & 90.58 \\
         Avg. Compression Ratio & 0.5 & 12.16 & 15.59 \\
         NHR only Compression Ratio & 5.81 & 5.81 & 5.81 \\
         Number of Images &
         4,326,979 &
         4,326,979 &
         4,326,979 \\
         Number of Hits & 744,150 & 744,150 & 744,150 \\ 
         Number Indexed & 255,065 & 255,918 & 255,385 \\ 
         Rsplit $\downarrow$ & 7.58\% & 7.08\% & 6.41\% \\ 
         CC1/2 $\uparrow$ & 0.997 & 0.997  & 0.997 \\ 
         CCano  $\uparrow$ & 0.087 & 0.104 & 0.085 \\ 
         Rwork  $\downarrow$ & 0.206 & 0.199 & 0.469 \\ 
         Rfree  $\downarrow$ & 0.231 & 0.223  & 0.498 \\ 
         Map-Model CC $\uparrow$ & 0.81 & 0.80 & 0.45 \\ 
    \bottomrule
    \end{tabularx}
\end{table}

\section{Performance Assessment}\label{sec:performanceassesment}

In this section we conduct an extensive performance analysis to evaluate the impacts performance optimizations.  We first consider the impacts of autotuning the use of threads in the compression experiment. Next we consider the scalability of our approach through the specific example of ROIBIN-SZ. We then use our tuned configurations to consider the overall throughput and compression ratios.  

\subsection{Autotuning Speedup}

First we consider the impacts of the autotuning approach described in Section~\ref{sec:performanceopt} for all compressors that have parallel components.
We present our results in Table~\ref{tab:speedup}.
In this table the untuned configuration allocates all cores in the machines to the compression task (how compression is currently used at LCLS in HDF5 files).
For chunk size = 1, there are not  many meaningful differences in the speedup,  largely due to insufficient work to overcome the thread start up costs.
Hence, we start our presentation with chunk size = 16, the next size considered.
Similar, results exist for chunksize = 32 and 64, the other chunksizes we considered.

The tuned allocation column indicates the allocation chosen by OptZConfig.
In this column \texttt{tasks} represents allocation of MPI ranks to cores, \texttt{blosc} threads to the blosc implementation, \texttt{roi} threads to the region of interest implementation, and \texttt{bin} threads to the binning implementation.
Each term appears in each row where it is an option.

We observe that OptZConfig chooses a configuration that
(1) does not allocate all threads to tasks,  indicating there are benefits to these parallel optimizations we preform and
(2)  differs from compressor to compressor, indicating automation was useful in that these would have been to have been determined  manually otherwise.
We note that  the OptZConfig often oversubscribes threads to the machine rather than using a single thread per core design.  Consider the \projectName{} configuration as an example. Each of the 30 tasks spawns 4 threads when it performs the binning operation.  If executed all in parallel, this would be 120 threads on a 40-core machine.  In practice, however, these threads do not always overlap because the serial portions of execution vary in their runtime, leaving underutilized cores that other processes can utilize.

We  further note that there are almost always benefits (although in some cases modest) to tuning the thread configuration for a particular compressor and system.  The next three columns of the  table show the compression throughput and speedup on a single node.  We consider scalability beyond a single node in the next section.
Depending on the configuration we see anywhere from a 2\% speedup to a 31\% speedup with median of 8.6\% increase in bandwidth.

The autotuning process for \projectName{} to process one file the lysozyme dataset took 15 seconds on 8 nodes (vs 4 seconds on 1 node for 1 tuned execution), but it then was able to be used for the remaining configurations and experiments.
This one-time setup cost would pay itself back quickly upon multiple invocations of the compressor.

\begin{table}[t]
\centering
\caption{Configurations with threads to allocate with meaningful change in performance on lysozyme, chunk size=16, 40 cores}
\label{tab:speedup}
\begin{tabularx}{\columnwidth}{llXXX}
\toprule
Configuration & Tuned Allocation &  Speedup &  Untuned &  Tuned \\
 &  &   \% &  GB/s &  GB/s \\
\midrule
blosc-6         & tasks=30,blosc=6             &  2.53 &          1.14 &           1.17 \\
blosc-9         & tasks=30,blosc=7             & 25.36 &          0.17 &           0.21 \\
blosc           & tasks=30,blosc=8             & 31.27 &          3.20 &           4.20 \\
roibin\_blosc-6 & tasks=30,roi=4,bin=4,blosc=1 & 11.40 &          3.58 &           3.99 \\
roibin\_blosc-9 & tasks=30,roi=4,bin=4,blosc=1 & 13.25 &          0.54 &           0.61 \\
roibin\_bzip-6  & tasks=30,roi=6,bin=6         &  5.25 &          0.91 &           0.96 \\
roibin\_bzip-9  & tasks=30,roi=6,bin=6         &  4.21 &          0.86 &           0.90 \\
roibin\_mgard   & tasks=30,roi=1,bin=4         &  9.42 &          0.79 &           0.86 \\
roibin\_sz3     & tasks=30,roi=1,bin=4         &  7.77 &          2.28 &           2.46 \\
roibin\_sz      & tasks=30,roi=1,bin=4         &  7.61 &         11.45 &          12.32 \\
roibin\_zfp     & tasks=30,roi=1,bin=4         &  8.64 &         11.82 &          12.84 \\
\bottomrule
\end{tabularx}
\end{table}

\subsection{Scalability}

Next we consider scaling to multiple nodes in Table~\ref{tab:scaleability} for ROIBIN-SZ.
For this experiment, we run the execution on an increasing number of tasks and cores that span multiple nodes.
We reuse the autotuned configurations from the preceding section.

As we scale from 30 to 300 tasks (40 to 400 cores), we find that the execution scales linearly from 7.56 GB/s to 70.69 GB/s.  We see similar performance for Se-SAD at scale with 400 cores at 60 GB/s in the next sections of Table~\ref{table:roibin:performance:both}.  Dataset-to-dataset variation in compressor bandwidth is to be expected with compressors such as SZ \cite{sz16,sz3}
This indicates that $\approx 67 \sim 84$ to  nodes would be required using this configuration to scale to the desired 500 GB/s bandwidth after NHR.

\begin{table}[tbh]
    \centering
    \caption{Scalability of ROIBIN-SZ for 1 file of lysozyme}
    \label{tab:scaleability}
    \begin{tabularx}{\columnwidth}{lXr}
    \toprule
       Ranks&Cores  & Compression Bandwidth \\
    \midrule
       30&40  & 7.56 \\ 
       60&80  & 11.88 \\ 
       120&160  & 14.80 \\ 
       150&200  & 19.89 \\ 
       300&400  & 70.69 \\ 
    \bottomrule
    \end{tabularx}
\end{table}

\subsection{Compression Ratios and Throughput at Scale}

Next we consider throughput and compression ratios at scale for the two datasets in Table~\ref{table:roibin:performance:both}.

We find that ROIBIN-SZ is able to achieve very high compression ratios, up to 196.5 relative to the uint16 raw data format from the detectors and able to meet the 500 GB/s bandwidth requirement with $\approx71-82$ nodes.
This is a substantial improvement over lossless methods that, relative to the detector's raw format, actually result in an increase in storage used.

The ROIBIN lossless methods do provide a benefit over simple lossless compression; however, the majority of this benefit comes from binning and not lossless compression.
In fact, the compression ratio from lossless compression of the binned data is lower than the compression ratio of the data as a whole.
This means that unlike the lossy compressors that we see in the next section, lossless compression performance is reduced rather than improved by binning.

We also find that the variants of our method, such as ROIBIN-MGARD and ROIBIN-ZFP, outperform their non-ROIBIN counter parts.  This result can be attributed to the relative speed of binning to these compressors algorithms and the comparative smoothness of the data after binning.
ROIBIN-ZFP takes the position as the highest compression bandwidth on both datasets, although this comes at a large trade-off in compression ratios.
With ROIBIN-ZFP, the 500 GB/s target could be met with $\approx 55 \sim 73$ nodes, respectively.

% \begin{table}
% \centering
% \caption{Compression Ratios (CR) and Bandwidth for Configurations that Pass Quality Assessment for lysozyme - file 1 on 300 Ranks / 400 Cores}
% \label{table:roibin:performance:lysozyme}
% \begin{tabularx}{\columnwidth}{lXr}
% \toprule
%         &  &  compression \\
%         config &  CR &  GBps \\
% \midrule
%       bzip2-9 &     0.54 &    1.72 \\
%        bzip2-6 &    0.54 &    1.88 \\
%          blosc &    0.58 &   24.35 \\
%        blosc-6 &    0.60 &    5.73 \\
%        blosc-9 &    0.60 &    1.07 \\
%  roibin\_bzip-6 &   2.14 &    7.00 \\
%  roibin\_bzip-9 &   2.15 &    6.55 \\
%   roibin\_fpzip &   2.22 &   36.41 \\
%   roibin\_blosc &   2.36 &   71.39 \\
% roibin\_blosc-6 &   2.40 &   21.74 \\
% roibin\_blosc-9 &   2.41 &    4.13 \\
%          mgard &    5.89 &    1.21 \\
%            zfp &   11.44 &   38.32 \\
%   roibin\_mgard &  18.42 &    6.10 \\
%     roibin\_zfp &  35.08 &   89.89 \\
%      roibin\_sz & 194.86 &   66.80 \\
%     roibin\_sz3 & 196.60 &   70.69 \\
% \bottomrule
% \end{tabularx}
% \end{table}

\begin{table}
\centering
\caption{Compression Ratios  and Bandwidth for Configurations That Pass Quality Assessment for Lysozyme and Selenobiotinyl-Streptavidin -- File 1 on 300 Ranks / 400 Cores}
\label{table:roibin:performance:both}
\begin{tabularx}{\columnwidth}{X|rr|rr}
\toprule
        &  &  Lysozyme & & Se-SAD \\
        &  &  Compression & & Compression\\
        config &  CR &  GB/s & CR & GB/s\\
\midrule
      bzip2-9 &     0.54 &    1.72 & 0.60      &  2.00   \\
       bzip2-6 &    0.54 &    1.88 & 0.60      &  2.18   \\
         blosc &    0.58 &   24.35 & 0.61    &   25.63 \\
       blosc-6 &    0.60 &    5.73 & 0.63     &  6.68   \\
       blosc-9 &    0.60 &    1.07 & 0.64     &  1.10   \\
 roibin\_bzip-6 &   2.14 &    7.00 & 2.20     &  6.91   \\
 roibin\_bzip-9 &   2.15 &    6.55 & 2.19     &  6.64   \\
  roibin\_fpzip &   2.22 &   36.41 & 2.45     &  35.81  \\
  roibin\_blosc &   2.36 &   71.39 & 2.46     &  76.35 \\
roibin\_blosc-6 &   2.40 &   21.74 & 2.50    &  24.85  \\
roibin\_blosc-9 &   2.41 &    4.13 & 2.51   &   4.26 \\
         mgard &    5.89 &    1.21 & 1.44     & 1.89    \\
           zfp &   11.44 &   38.32 & 1.81     & 26.46  \\
  roibin\_mgard &  18.42 &    6.10 & 7.70     & 7.09    \\
    roibin\_zfp &  35.08 &   \textbf{89.89} & 10.52   & \textbf{66.33}   \\
     roibin\_sz & 194.86 &   66.80 &  \textbf{46.44}  &  50.61  \\
    roibin\_sz3 & \textbf{196.60} &   70.69 &  39.10   & 60.27 \\
\bottomrule
\end{tabularx}
\end{table}

% \begin{table}
% \centering
% \caption{Compression Ratios (CR) and Bandwidth for selenobiotinyl-streptavidin - file 1 on 300 Ranks / 400 Cores}
% \label{tab:roibin:performance:sadse}
% \begin{tabularx}{\columnwidth}{Xrr}
% \toprule
%         config &  CR &  compress\_bandwidth\_GBps \\
% \midrule
%       bzip2-6 &     0.60 &                  2.18 \\
%        bzip2-9 &    0.60 &                  2.00 \\
%          blosc &    0.61 &                 25.63 \\
%        blosc-6 &    0.63 &                  6.68 \\
%        blosc-9 &    0.64 &                  1.10 \\
%          mgard &    1.44 &                  1.89 \\
%            zfp &    1.81 &                 26.46 \\
%  roibin\_bzip-9 &   2.19 &                  6.64 \\
%  roibin\_bzip-6 &   2.20 &                  6.91 \\
%   roibin\_fpzip &   2.45 &                 35.81 \\
%   roibin\_blosc &   2.46 &                 76.35 \\
% roibin\_blosc-6 &   2.50 &                 24.85 \\
% roibin\_blosc-9 &   2.51 &                  4.26 \\
%   roibin\_mgard &   7.70 &                  7.09 \\
%     roibin\_zfp &  10.52 &                 66.33 \\
%     roibin\_sz3 &  39.10 &                 60.27 \\
%      roibin\_sz &  46.44 &                 50.61 \\
% \bottomrule
% \end{tabularx}
% \end{table}

\section{Discussion and Conclusions}\label{sec:conclusions}

In this work we have shown substantial improvements to the compression of serial crystallography data while performing science-preserving compression.
We were able to achieve up to a 196$\times$ and 46.44$\times$ compression ratio (and up to 631$\times$ and 154.18$\times$ total when non-hit rejection is used additionally) respectively on lysozyme and selenobiotinyl-streptavidin while preserving the data sufficiently to reconstruct the structure at bandwidths and scales that are reasonably obtainable to match the compression bandwidth needs of the upcoming APS-II and LCLS-II-HE systems.

For future work, we would like to consider the application of this approach to other domains.
Additionally, we would like to consider other forms of light source data that may benefit from our approach and experiment with FPGAs/GPUs. %\robert{Chuck any other obvious crystallography future work}

%  REMOVED FOR DOUBLE BLIND REVIEW
%  \section*{Acknowledgment}
%  
%  This research was supported by the Exascale Computing Project
%  (17-SC-20-SC), a collaborative effort of the U.S. Department
%  of Energy Office of Science and the National Nuclear Security
%  Administration. Use of the Linac Coherent Light Source (LCLS),
%  SLAC National Accelerator Laboratory, is supported by the U.S.
%  Department of Energy, Office of Science, Office of Basic Energy
%  Sciences under Contract No. DE-AC02-76SF00515.
%  This material is based upon work supported by the U.S. Department of Energy, Office of Science, under contract number DE-AC02-06CH11357. 
%  
\bibliography{roibinsz}
\bibliographystyle{IEEEtran}

\end{document}